\documentclass[%
preprint,
 amsmath,amssymb,
pra,
]{revtex4-1}

\usepackage{mathrsfs}
\usepackage{graphicx}
\usepackage{hyperref}
\begin{document}

\title{The time geography of segregation during working hours}


\author{Teodoro Dannemann}
\author{Boris Sotomayor-G\'omez}
\author{Horacio Samaniego}
\email[Corresponding author: ]{horacio@ecoinformatica.cl}

\affiliation{Ecoinform\'atica Lab, Universidad Austral de Chile, Campus Isla Teja s/n, Valdivia, Los R\'ios, Chile}



\date{\today}
\begin{abstract}
Understanding segregation is essential to develop planning tools for building more inclusive cities. Theoretically, segregation at the work place has been described as lower compared to residential segregation given the importance of skill complementarity among other productive factors shaping the economies of cities. This paper tackles segregation during working hours from a dynamical perspective by focusing on the movement of urbanites across the city. In contrast to measuring residential patterns of segregation, we used mobile phone data to infer home-work trajectory networks and apply a community detection algorithm to the example city of Santiago, Chile. We then describe qualitatively and quantitatively outlined communities, in terms of their socio economic composition. We then evaluate segregation for each of these communities as the probability that a person from a specific community will interact with a co-worker from the same community. Finally, we compare these results with simulations where a new work location is set for each real user, following the empirical probability distributions of home-work distances and angles of direction for each community. Methodologically, this study shows that segregation during working hours for Santiago is unexpectedly high for most of the city with the exception of its central and business district. In fact, the only community that is not statistically segregated corresponds to the downtown area of Santiago, described as a zone of encounter and integration across the city.
\end{abstract}


\maketitle

\section{Introduction}

The historical and unprecedented growth of income inequality worldwide has pushed segregation to a pivotal concept in the description of social systems \cite{grusky}. For instance, segregation is known to generate far reaching impacts for individuals and groups by altering opportunities for education, employment, health care, and general well fare, among others. Segregation has extensively been described to effect societal imbalances leading to critical states in terms of security, health and wealth distribution \cite{silm2014temporal,cutler1997ghettos,massey1987effect,Garreton2016a}, while social cohesion has been posed as a fundamental process fostering the enrichment of the social capital available at particular locations \cite{farber2013social, coleman1988social, forrest2001social}. However, defining and measuring segregation remains a complex and elusive task in which scientists have recognized several dimensions, which are not integrated in a general conceptual framework \cite{louf2016patterns}. So far, standard methods have mostly relied on static views seeking to describe and characterize residential ghettoization  \cite{yip2016exploring,wong2011measuring} and several indices have been put forward to quantify inequality across residential areas \cite{jones2014redefining}, being the most paradigmatic example the  Duncan Dissimilarity Index \cite{duncan1955methodological}, which measures the percentage of minority population that would have to be relocated in order to perfectly integrate among the distribution of residents of a region \cite{farber2015measuring}. However, these indices do not often consider social interactions in other contexts such as work and leisure to define segregation. Such vertical view of social integration seem thus to be a fundamental aspect that, while embedded in the concept of segregation, is not often considered for the study of the geography of inequality. The explicit consideration of this latter approach has led to new insights for the study of segregation that mostly focus on the so-called \textit{exposure dimension} of segregation in an attempt to capture ``the extent in which members of one group encounter members of another group in their local spatial environments'' \cite{reardon2004measures,massey1988dimensions}. Hence, the concept of exposure explicitly takes into account the set of spaces that every person visits during his daily journey -also called the activity space within the subdiscipline of time geography \cite{thrift1977,Hagerstraand2005}. For instance, Wong and Shaw \cite{wong2011measuring} used daily travel data surveys in conjunction with racial-ethnic data to calculate the exposure (or, conversely, the isolation) level of different ethnic groups in southeast Florida. Similarly, Farber et al \cite{farber2015measuring} used the time-geography framework and origin-destination surveys to estimate the Social Interaction Potential index, given by the spatiotemporal prism generated between all possible paths between home and work. 

Nowadays, the explosive use of communication technologies, such as cellphones, have made huge volumes of non conventional data available for research purposes. For instance, by knowing to which cellphone tower we connect across the day permits the reconstruction of peoples daily  trajectories, hence providing a surprisingly high spatio-temporal resolution of social interactions across space \cite{panigutti2017assessing} . This approach has been widely used recently to assess a variety of topics going from individual mobility patterns \cite{gonzalez2008understanding} and land use patterns \cite{lenormand2015comparing}, to the detection of relevant places of high social activity within the city \cite{grauwin2015towards}, thereby unveiling the structure of cities \cite{louail2014mobile,Lenormand2015f}. Ratti et al \cite{ratti2010redrawing} used the Newman's community detection algorithm \cite{newman2006modularity} on Call Detail Records (CDR) to divide Great Britain into zones having large internal interactions (i.e. call volumes) as compared to interactions with other zones. Interestingly, these automatically-detected communities show high correlation with administrative regions of Great Britain. Since then, several studies of mobility networks have been published unveiling meaningful communities out of functional social activities at the level of individual cities, regions, and countries \cite{gao2017uncovering,sobolevsky2013delineating,zhong2013identifying}. However, some concern has been voiced given the simplicity of the null model used in Newman's algorithm (i.e. the Erdos Renyi network, a purely random network) as detected communities could simply be a consequence of the local movement of individuals. More realistic approaches have since emerged to include gravitational effects within the null model \cite{sarzynska2015null,expert2011uncovering}.\par 

In this contribution, we combine social segregation (in particular social isolation) and community detection algorithms to provide a robust tool-set for describing urban segregation and, particularly, its exposure dimension. We use the isolation index $P_\mathscr C$ \cite{massey1988dimensions} as a measure to evaluate the efficiency of Newman's modularity method for finding socially segregated communities within the city. We accomplish this by creating random walk simulations and the network generated by them as a null model, comparing the real values of $P_\mathscr C$ to simulation outputs, and finding significant deviations from null expectations in our case study city (Santiago, Chile). We also use censal data to evaluate correlations between communities and socioeconomic level. 

\section{Materials and Methods} \label{MM}
\subsection{Dataset}
We use anonymized CDR data of mobile phone users at a spatial resolution of individual cellphone towers. Data was provided by Telef\'onica Chile and represents a 37\% share of the mobile phone market in Chile. The dataset consists of all cellphone pings to an antenna along four working weeks (from Monday to Friday) in March, May, October, and November 2015, summing a total of $9\times 10^8$ data records and $3.5\times10^5$ individual users across Santiago, Chile.
Only cellphone towers within the urban boundary of Santiago were considered based on the official administrative registries \cite{MINVU}. Voronoi tesselations were constructed around each tower to represent its spatial coverage area. In order to discard rural areas, only cellphone towers with a minimum of 70\% overlap between its Voronoi area and the urban area were considered.

\subsection{Home and work place definition}
\label{HWdef}
Following  Phithakkitnukoon et al \cite{phithakkitnukoon2012socio} and Scepanovic et al \cite{vscepanovic2015mobile}, among others, we inferred the home and work location for each user when possible. Each home location was labeled as the most frequented (i.e. pinged) place (tower) between 10pm and 7am. Likewise, work location was defined as the place with more pings between 9am and 5pm. We also required each user to have, at least, five connection to a potential home or work place, and to additionally have  $> 50\%$ of their pings made from home or work. If any of these criteria were not fulfilled, data records were discarded. 

\subsection{Network construction and community detection}
Undirected weighted networks were built based on H-W trajectories per user. Nodes represent towers and weighted links the number of H-W trajectories shared by two towers. Blondel's community detection algorithm \cite{blondel2008fast} was implemented to spatially segment towers in the network (and their corresponding Voronoi areas). This algorithm is based --as many detection algorithms do-- on the maximization of the network modularity, which measures the density of links inside each community as compared to links between communities \cite{newman2006modularity,girvan2002community,blondel2008fast}. The method proposed by Blondel et al \cite{blondel2008fast} has shown to be of high performance in terms of accuracy and computing time compared to other classical methods \cite{yang2016comparative}.
Even after filtering, some antennas remained without affiliation to any of the detected communities forming interspersed communities of individual antennas.  We discarded those nodes and labeled each user with the community corresponding to its home location. 

The main network corresponds to the one generated for all four weeks of data. However, we additionally generated a network for each working day of the week, and compared the communities generated in each one with respect to the main (aggregated) network.

\subsection{Socioeconomic composition of communities}
The classification of socioeconomic level (SEL) is taken from ADIMARK \cite{adimark2009mapa} (figure \ref{mapastgo}a), which defines SEL from the national census data taking into account two dimensions: educational level and the ownership of material assets (see Supplementary Material 1 for further methodological details). ADIMARK identifies five relevant groups labeled: S1, S2, S3, S4, and S5, with S1 as the most affluent group and S5 the group with the lowest income and educational achievements. Spatially, each census block is assigned to one SEL group. Because blocks are variable in size and shape, SEL was assigned to a Voronoi cell by weighing SEL of each censal block by the areal contribution to each cell (see Supplementary Material 2 for further details). We then aggregated all Voronoi cells corresponding to a specific community to obtain the final SEL composition of each community.

\subsection{Isolation index as a measure of segregation} \label{II}
As Massey and Denton \cite{massey1988dimensions} proposed in their seminal work, isolation of a certain group (e.g. community) $\mathscr C$ can be measured as:
\begin{equation}
P_\mathscr C=\sum_{i=1}^{n} \frac{c_i}{C}  \frac{c_i}{T_i}
\label{eq1}
\end{equation}                                           
where $c_i$ is the number of cellphone users of community $\mathscr C$ in area unit $i$, $C$ is the total number of users belonging to community $\mathscr C$, and $T_i$ the total count of users in $i$. In our case, each user will belong to the community its home belongs, and we calculated isolation index in workplace, i.e. the areal unit $i$ denotes workplaces. Consequently, $\frac{c_i}{C}$ denotes the probability that a member of (whose home belongs to) community $\mathscr C$ will work in unit $i$ and $\frac{c_i}{T_i}$ is the fraction of users belonging to community $\mathscr C$ working in unit $i$. Hence, $P_{\mathscr C}$ is the probability that a user of community $\mathscr C$ will randomly interact with someone of its same community at its work location $i$, making Eq. \ref{eq1} a direct measure of the level of isolation with members of your own community ($\mathscr C$) while at work. Thus, large $P_{\mathscr C}$ may denote highly segregated communities, while smaller values are indicative of more integrated ones. As defined here,  $P_\mathscr C$ is bounded between two limit cases: The first case may be thought of as the "well-mixed limit", i.e., if for each user, work location was chosen completely at random and therefore members of all communities mingle. Then, $\frac{c_i}{T_i}=k$, with  $k$ the proportion of group $\mathscr C$ to the total population. Replacing this in equation \ref{eq1} we obtain $P_\mathscr C=k$, showing that, when the population is completely mixed, the probability of encounter between members of the same community is merely the proportion of this community to the total population size. In the other limit the population is completely segregated (i.e. isolated) and cellphone users belonging to a particular community will share their workplace only with members of their own community $\mathscr C$ (i.e. $\frac{c_i}{T_i}=1$), and one easily gets $P_\mathscr C \rightarrow 1$. We calculated real isolation indexes (\emph{RII}) for each community from our dataset, and a simulated isolation index (\emph{SII}) from simulations performed as follows. 

\subsection{Random walks and robustness of community detection}
Using a randomization procedure we evaluated how likely is to obtain the real H-W trajectories and their respective isolation indices. To such aim, we characterized the displacement of users by recovering the statistical distribution of (i) H-W distances ($D_{HW}$) and (ii) angles of direction ($\theta$) of H-W journeys  (with respect to East direction) for each community. We then  randomly draw $D_{HW}$ and $\theta$  for each user from the empirical distribution of its community to obtain a new simulated work location. From this, and by maintaining the original home location  (i.e. Voronoi cell) we obtained new simulated H-W trajectories to compute the simulated isolation index (\textit{SII}) and compare it against \textit{RII}. Further details on these simulations are included in Supplemental Material 4.

\section{Results}
\subsection{Description of communities}
\begin{figure}[h]
  \raggedleft
      \includegraphics[scale=0.5]{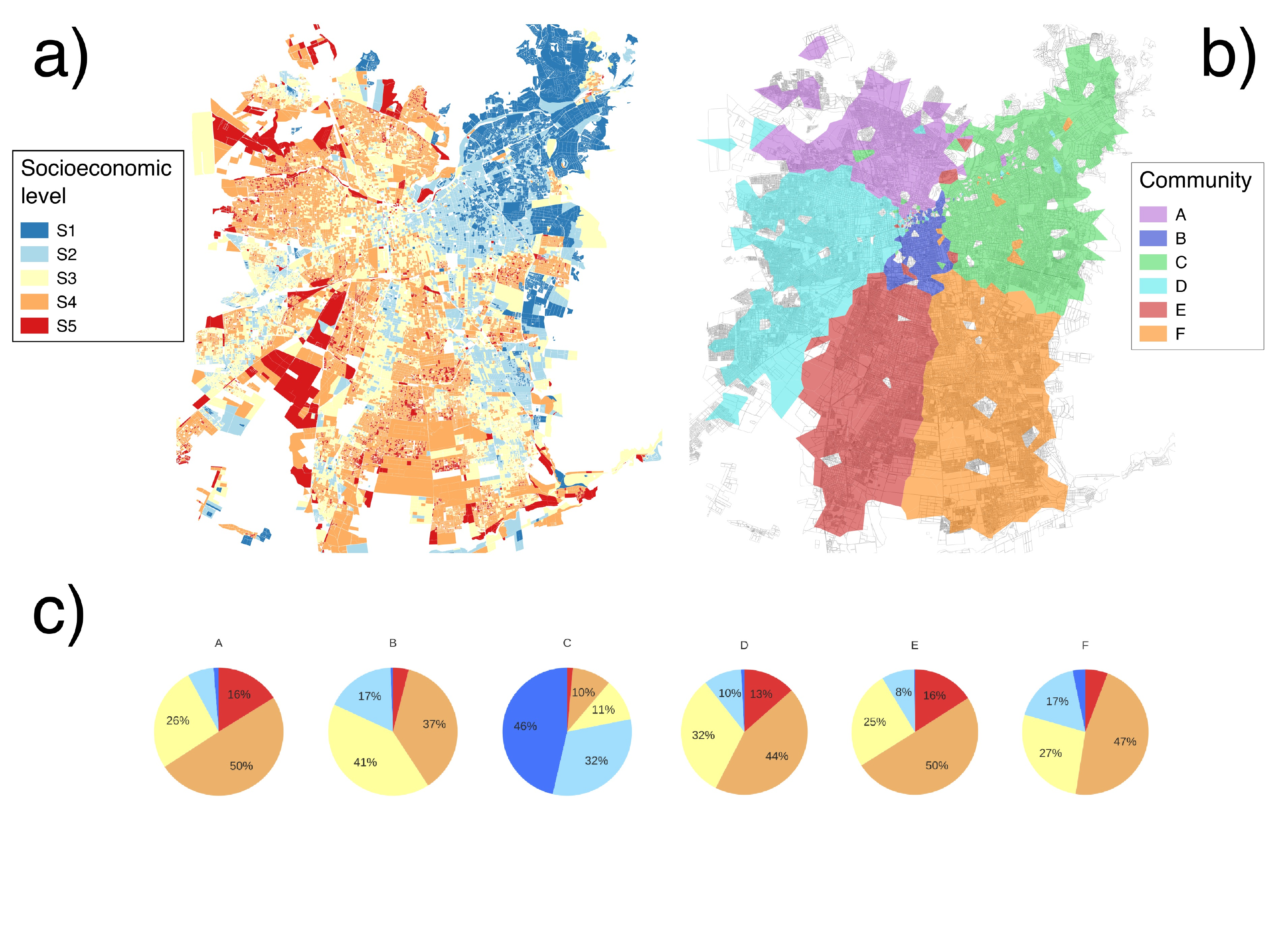}
  \caption{Santiago, Chile. a) Spatial distribution of socioeconomic level \cite{}. b) Spatial distribution of detected communities using the Blondel's algorithm. c) Socioeconomic level composition of each detected community .}
  \label{mapastgo}
\end{figure}

Six communities were retrieved from Santiago's H-W aggregated network (figure \ref{mapastgo}b). Notably, daily networks were also split into six communities highly consistent with communities on the aggregated network (see Supplementary Material 3 with detailed community changes across daily networks). Table \ref{tab:Sens} shows the percentage of nodes (cellphone towers) for each weekday  that retains its community affiliation in the aggregated network. At least 75\% of such nodes retain their  community affiliation, independent of the week day chosen.

A qualitative comparison of figures \ref{mapastgo}a and \ref{mapastgo}b shows an intriguing correspondence between the distribution of SEL  and detected communities. Figure \ref{mapastgo}c complements such view by highlighting the specific SEL composition of each detected community. Community $C$ has, by far, the highest fraction of most affluent SEL ($S1$ and $S2$). These groups, however, constitute less than a 10\% of communities $A$, $D$, and $E$, where $S3$ and $S4$ dominates.  Community $F$ resembles  $A$, $D$ and $E$, but has a larger presence of more affluent users. Finally, group $B$ is in between extremes, and  has a high composition of middle SEL ($S3$).

\begin{figure}
  \centering
      \includegraphics[width=1.0\textwidth]{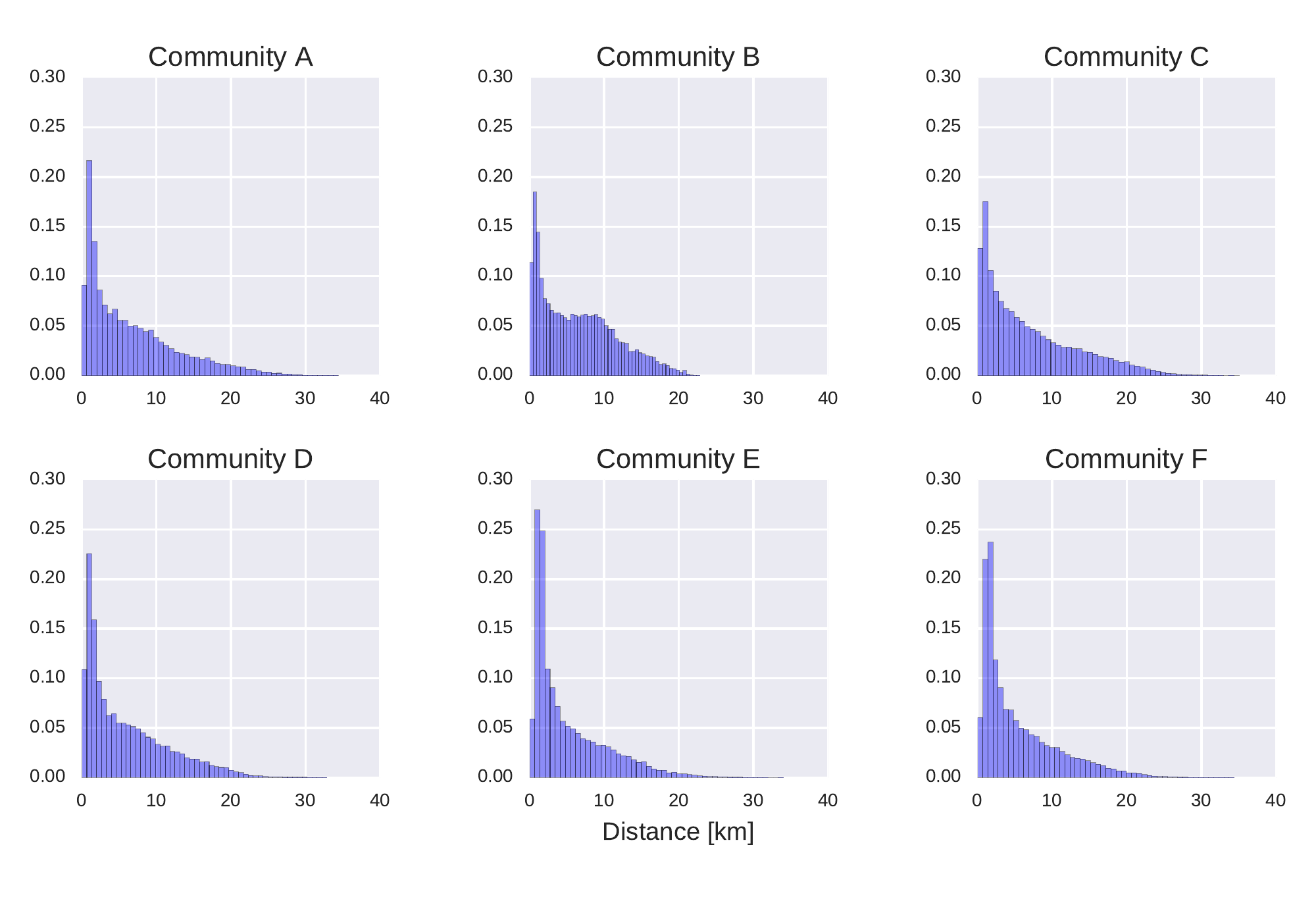}
  \caption{Probability distribution of traveled distance of users affiliated to  each of the six communities detected in Santiago, Chile.}
 \label{histdist}
\end{figure}

\begin{figure}
  \centering
      \includegraphics[width=1.0\textwidth]{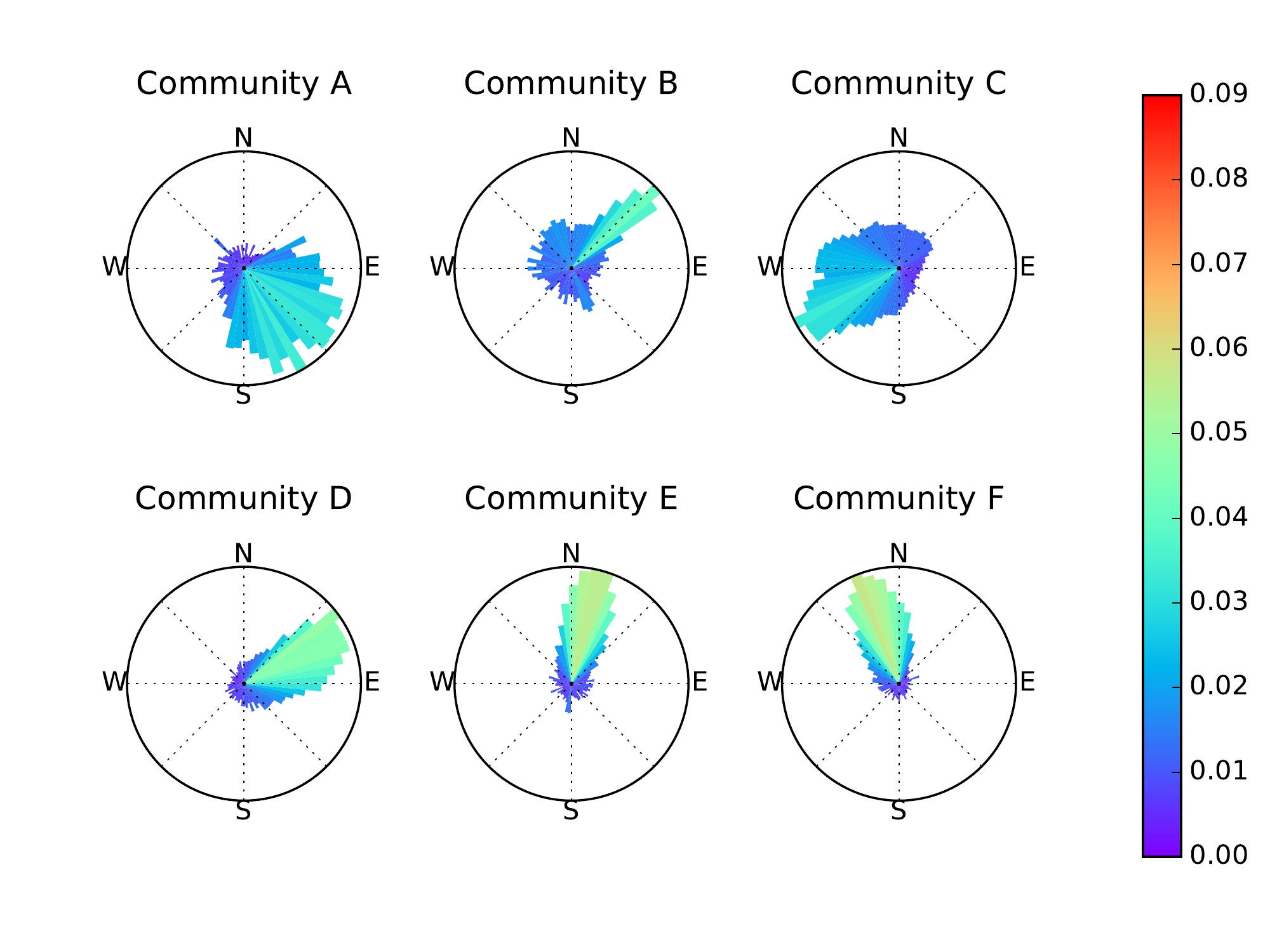}
  \caption{Probability distribution of  H-W movement angles across individuals belonging to each of the six communities detected in Santiago, Chile. Colorbar shows the probability density estimation in the direction angle.}
  \label{histang}
\end{figure}

H-W distances (figures \ref{histdist}) show a mean journey of  5 to 7 kilometers among communities (table \ref{DIST}).  Similarly, the distribution of H-W angles (figure \ref{histang}) suggests a radial movement configuration pointing to the center of the city (community $B$), which seems to attract movement vectors of each community to it. 

\begin{table}
  \begin{tabular}{c|ccccc}
      & \textbf{Monday} & \textbf{Tuesday} & \textbf{Wednesday} & \textbf{Thursday} & \textbf{Friday}\\
      \hline
      $\textbf{Retained nodes}$ & 81.21\% & 77.53\% & 80.59\% & 75.26\% & 79.28\% \\[1mm]
  \end{tabular}
  \caption{\label{tab:Sens} Percentage of nodes retaining  community affiliation, compared to the aggregated network.}
\end{table}

\begin{table}
  \begin{tabular}{c|cccccc}
      & \textbf{A} & \textbf{B} & \textbf{C} & \textbf{D} & \textbf{E} & \textbf{F} \\ 
      \hline
      $\textbf{Mean (km)}$ & 6.65 & 6.51 & 6.97 & 6.05 & 5.19 & 5.50 \\[1mm]
      $\textbf{Standard Deviation (km)}$ & 6.02 & 5.03 & 6.20 & 5.45 & 4.96 & 5.15 \\[1mm]
  \end{tabular}
  \caption{\label{DIST} Mean and standard deviation of distance traveled by community.}
\end{table}

\subsection{Isolation and H-W segregation}
\begin{figure}[h!]
  \centering
      \includegraphics[width=0.8\textwidth]{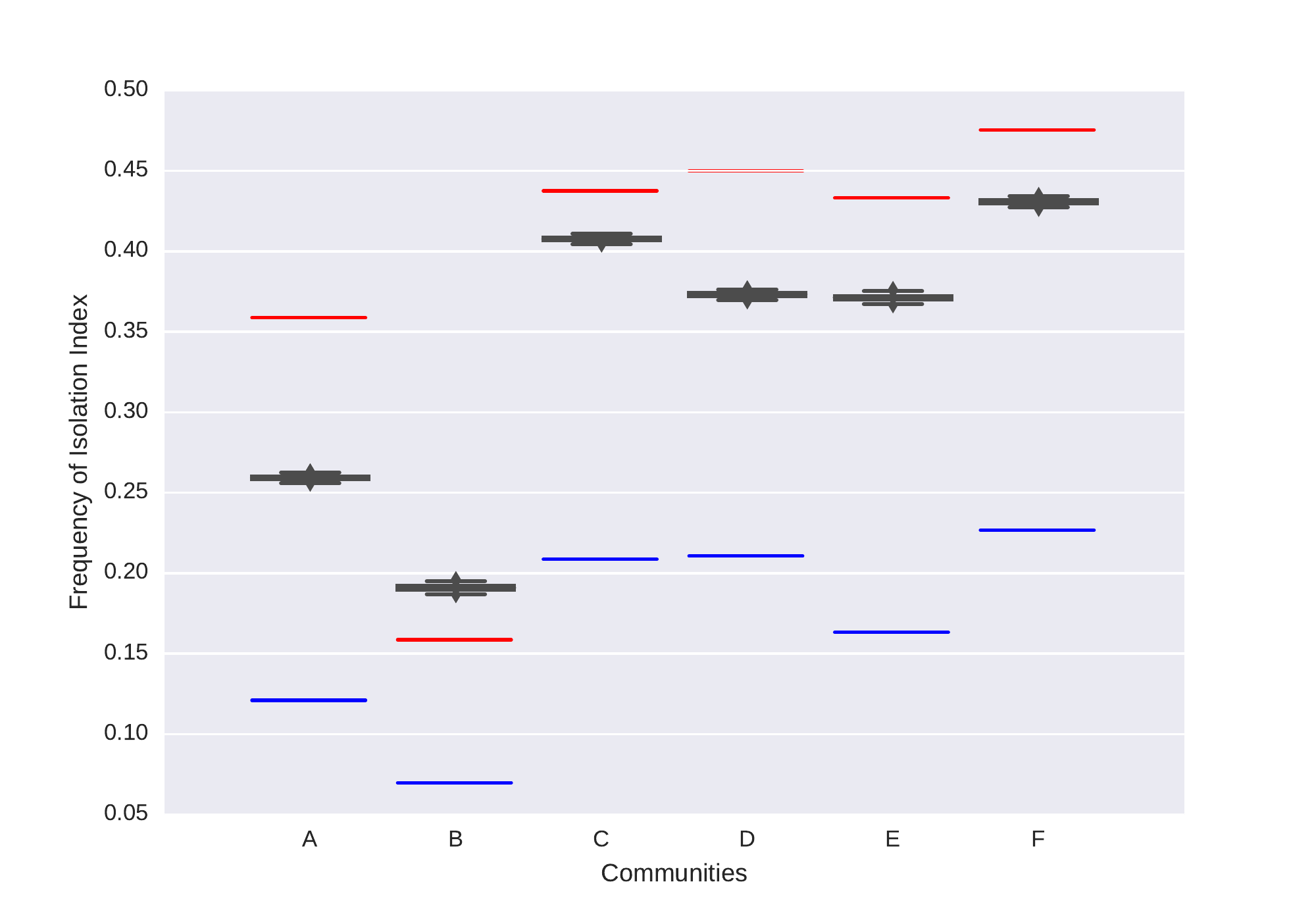}
  \caption{Isolation index value for each community. Real isolation index ($RII$) values are indicated in red lines, while values obtained from simulations ($SII$) are depicted by black boxplots. Blue lines show isolation index values in the ``well mixed limit'', as explained in section \ref{MM}}
 \label{isolation}
\end{figure}

The comparison of \textit{RII} to \textit{SII}, in figure \ref{isolation}, shows that five out of six communities have real values (red segments) much larger than expected based on our simulation framework. Community $B$ seem to be the only exception and shows smaller \textit{RII} values ($0.159$)  compared to randomized  values ($SII=0.191$) suggesting  this region  as the only one that is not statistically segregated (downtown Santiago, mainly). Specific isolation index values are shown in table \ref{Sim}. The distance of $RII$ from $SII$, in terms of the standard deviation of simulations ($\sigma_{SII}$) shows the likelihood to obtain \textit{RII} from our simulation. We also  show in blue segments in figure \ref{isolation} the isolation index value in the hypothetical ``well-mixed limit'' ($WII$), that is, drawing work places not from the known $\theta$ and $D_{HW}$, but from a completely random uniform  distribution independent from their home location.

\begin{table}[h!]
  \begin{tabular}{c|cccccc}
     \textbf{Value} & \textbf{A} & \textbf{B} & \textbf{C} & \textbf{D} & \textbf{E} & \textbf{F} \\
      \hline
      $\textbf{RII}$                 & 0.359 & 0.159 & 0.438 & 0.450 & 0.433 & 0.476 \\[1mm]
      \boldmath$\langle SII \rangle$ & 0.259 & 0.191 & 0.408 & 0.373 & 0.371 & 0.431 \\[1mm]
      \boldmath{$\sigma_{SII}$}      & 0.0013 & 0.0016 & 0.0011 & 0.0012 & 0.0015 & 0.0013 \\[1.5mm] 
       \boldmath{$\frac{\mid RII-\langle SII \rangle \mid}{\sigma_{SII}}$} & 74.7 & 20.08 & 26.19 & 64.09 & 42.04 & 35.28 \\[3mm] 
  \end{tabular}
  \caption{ \label{Sim} Comparison of isolation indexes obtained by real data ($RII$, red segment in figure \ref{isolation}) and simulations ($SII$, black boxes in figure \ref{isolation}), in which case we show average value $\langle SII \rangle$ and standard deviation $\sigma_{SII}$. In the last row we calculated the separation of $RII$ with respect to $SII$ in  standard deviation units, $\sigma_{SII}$.}
\end{table}

\section{Discussion}

Understanding segregation has proven to be a multidimensional issue highly regarded by sociologist, economists, and social planners in general. We propose here a methodological framework that delves into the  exposure dimension of segregation  of urbanites while in their work location. We show that the modularity optimization and the community detection algorithm proposed by Blondel et al \cite{blondel2008fast} is an efficient method to outline communities representing groups that segregate through movement. We combine this with the use of an isolation index at the individual level to evaluate the degree of interaction between these groups during working hours. We highlight that communities found here are not simply given by the geographic context or the  urban transport network of our example city, but rather, they are driven by the bias resulting from the social interaction across members of their own communities. This is methodologically explained by the fact that our null (random walk) model retains the  statistical structure of urbanite's journey-to-work paths, i.e., it takes into account the well-known gravitational effect of spatial networks \cite{expert2011uncovering}.

It is interesting to note that, to our knowledge, the combination  of community detection algorithms with segregation tools is a novelty, and provides new insight to further the understanding of the complex geography of segregation during working hours. While the high isolation values obtained by our analysis confirm patterns of residential segregation already documented for Santiago \cite{caceres2017peri, agostini2016segregacion, Garreton2017}, our approach goes beyond static representations by inspecting the dynamic processes of social interactions while at work. Also, the high spatial and temporal resolution of mobile phone data allows for a much better understanding of the structure of the city, when compared to analyses from origin-destination or other conventional surveys \cite{lotero2016rich, fuentes2017santiago, farber2015measuring, wong2011measuring}. In particular, the communities found here confirm those sociological descriptions that (roughly) divide Santiago into a rich part in the foothills to the east, and a less affluent zone to the west and south \cite{caceres2017peri,agostini2016segregacion}. In our work, community $C$ closely matches the so-called ``high rent cone of Santiago'' \cite{link2015segregacion}. In fact,  the cone shaped community $C$ may be thought of as the footprint of the upper class zones described for Latin American cities in such literature \cite{sabatini2001segregacion}, where it has been proposed to emerge from the historical movement of the elites out of downtown areas (usually central) towards arbitrary radial directions in the periphery. This is further corroborated by inspecting the socioeconomic composition of community $C$ where the vast majority of SEL $S1$ and $S2$ is concentrated (figure \ref{mapastgo}c). Communities $A$, $D$ and $E$ instead, locate to the west, composing the less affluent social fabric of the city (see also their respective socioeconomic composition in figure \ref{mapastgo}c). Community $F$ contains the most densely populated boroughs (\emph{comunas}) of La Florida and Puente Alto, seen by some scholars as sub-central outgrows \cite{truffello2015policentrismo} where most of the middle class families live.  Finally, community $B$ represents the downtown Santiago, and it has a more diverse socioeconomic composition. 

Interestingly, community $B$ is the only one in which urbanites show no statistical segregation at their workplaces, as opposed to the other five communities where urbanites are likely to find co-workers belonging to their same community. These findings rise new concerns, especially if one takes into account the well accepted hypothesis that social segregation is lower at workplaces compared to residential places \cite{schnell2001sociospatial, ellis2004work, silm2014temporal}. Hence, our results emphasizes the importance of assessing segregation not only from a spatial and static point of view, but also using temporal assessments of segregation, such as proposed by Silm and Ahas \cite{silm2014temporal}, where, they assess temporal changes of ethnic exposure levels from mobile phone data across the day.

\section{Conclusion}
In this work we show that segregation and, particularly, the exposure dimension of segregation can be successfully combined with new technological tools such as big data generated from cellphone usage in order to explore the details of interaction patterns across the city. As a bridge between raw data (which is meaningless by its own) and sociological conclusions, there is a network approach and, particularly, community detection algorithms that provide added value to the data. Communities found here show high segregation patterns, that is, an average low exposure of an individual to other groups, when compared with random walk simulations. In our case, and for the sake of simplicity, we characterize daily journeys with only home and work locations for each person. However, the  increasing availability of rich cellphone data makes possible to have suitable proxies to generate detailed user trajectories that, when crossed with other socio demographic information, will certainly rise the possibility of advancing the understanding of  social interactions, segregation and other sociological topics in the urban area. 

In summary, the main results here are twofold: firstly, the presentation of a novel methodology that incorporates both community detection algorithms and segregation tools, in order to study segregation patterns across the urban area in a high spatio-temporal resolution. Secondly, we address the specific case of Santiago, and corroborate that this analysis gives useful information when combined with other sources of data, such as socioeconomic level or other census data, which may be used by urban planners, politicians, and social scientists in general. 

\bigskip
\noindent
\textbf{Data Accessibility:} Dataset containing user's home and and work locations as well as community affiliation of each tower are available from Dryad (\url{https://doi.org/10.5061/dryad.h8405}).

\bigskip
\noindent
\textbf{Author's Contributions:} T.D, B.S. and H.S. conceived the study. T.D., B.S. and H.S. extracted the data. T.D., B.S. and H.S. performed calculations. T.D., B.S. and H.S. wrote the manuscript.

\bigskip
\noindent
\textbf{Competing Interests:} Authors declare to have no competing interests.

\bigskip
\noindent
\textbf{Funding:} Funding for this research was provided by FONDEF-CONICYT grant \#ID15I10313 and  FONDECYT-CONICYT grant \#1161280 to H.S.

\bibliographystyle{vancouver}

\end{document}